\def\IEEEsubmission{0}
\if\IEEEsubmission1
\documentclass[journal,12pt,onecolumn,draftclsnofoot]{IEEEtran}
\else
\documentclass[journal]{IEEEtran}
\fi
\usepackage{float}
\usepackage[utf8]{inputenc}
\usepackage{multicol}
\usepackage{mathtools}
\usepackage{amsthm}

\usepackage{amsfonts}
\usepackage[bookmarksopen=true]{hyperref}
\usepackage{stfloats}
\usepackage{acro}
\usepackage[noadjust]{cite}
\usepackage{multirow}
\usepackage{float}
\usepackage{bm}
\usepackage{enumitem}
\usepackage{amsmath,amssymb}
\usepackage{graphicx}
\usepackage{epstopdf}
\usepackage[table,xcdraw]{xcolor}
\usepackage[geometry]{ifsym}
\usepackage{array}
\usepackage[utf8]{inputenc}
\usepackage[T1]{fontenc}
\usepackage{pifont}

\usepackage{algorithmic}
\usepackage{algorithm}
\usepackage[caption=false,font=normalsize,labelfont=sf,textfont=sf]{subfig}
\usepackage{textcomp}
\usepackage{url}
\usepackage{verbatim}
\usepackage{xcolor}
\usepackage[normalem]{ulem}
\providecolor{added}{rgb}{0,0,1}
\providecolor{deleted}{rgb}{1,0,0}

\def\BibTeX{{\rm B\kern-.05em{\sc i\kern-.025em b}\kern-.08em
		T\kern-.1667em\lower.7ex\hbox{E}\kern-.125emX}}

\DeclareAcronym{IoT}{short = IoT, long  = internet-of-things}
\DeclareAcronym{BC}{short = BC, long  = backscatter communication}
\DeclareAcronym{BS}{short = BS, long  = base station}
\DeclareAcronym{BD}{short = BD, long  = backscatter device}
\DeclareAcronym{SR}{  short = SR,  long  = symbiotic radio}
\DeclareAcronym{OFDM}{short = OFDM,long  = orthogonal frequency division multiplexing}
\DeclareAcronym{CP}{short = CP,long  = cyclic prefix}
\DeclareAcronym{ISAC}{short = ISAC,long  = integrated sensing and communication}
\DeclareAcronym{DLI}{short = DLI,long  = direct-link interference}
\DeclareAcronym{IBDI}{short = IBDI,long  = inter-backscatter device interference}
\DeclareAcronym{SNR}{short = SNR,long  = signal-to-noise ratio}
\DeclareAcronym{SINR}{short = SINR,long  = signal-to-interference-and-noise ratio}
\DeclareAcronym{SIC}{short = SIC, long = successive interference cancellation}
\DeclareAcronym{PAPR}{short = PAPR,long  = peak to average power ratio}
\DeclareAcronym{DFT}{short = DFT,long  = discrete Fourier transform}
\DeclareAcronym{3GPP}{ short = 3GPP, long = 3rd generation partnership project } 
\DeclareAcronym{IDFT}{short = IDFT,long  = inverse discrete Fourier transform}
\DeclareAcronym{FFT}{short = FFT,long  = fast Fourier transform}
\DeclareAcronym{IFFT}{short = IFFT,long  = inverse fast Fourier transform}
\DeclareAcronym{NMSE}{short = NMSE,long  = normalized mean square error}
\DeclareAcronym{AWGN}{short = AWGN,long  = additive white Gaussian noise}
\DeclareAcronym{BER}{  short = BER,  long  = bit error rate}
\DeclareAcronym{RMSE}{ short = RMSE,  long  = root mean squared error}
\DeclareAcronym{A-IoT}{ short = A-IoT,  long  = ambient IoT}
\DeclareAcronym{ISABC}{ short = ISABC,  long  = integrated sensing and backscatter communication }
\DeclareAcronym{RF}{short = RF, long = radio frequency}
\DeclareAcronym{SBC}{short = SBC, long = symbiotic backscatter communication}
\DeclareAcronym{DSK}{short = DSK, long = delay shift keying}
\DeclareAcronym{OOK}{short = OOK, long = on-off keying}
\DeclareAcronym{ICI}{short = ICI, long = inter-carrier interference}
\DeclareAcronym{RIS}{short = RIS, long = reconfigurable intelligent surfaces}
\DeclareAcronym{AFDM}{short = AFDM, long = affine frequency domain multiplexing}
\DeclareAcronym{Rx}{short = Rx, long = receiver}
\DeclareAcronym{IDAFT}{short = IDAFT, long = inverse discrete affine Fourier transform}
\DeclareAcronym{ISI}{short = ISI, long = intersymbol interference}
\DeclareAcronym{LTV}{short = LTV, long = linear time variant}
\DeclareAcronym{CSI}{short = CSI, long = channel state information}
\DeclareAcronym{SAW}{short = SAW, long = surface acoustic wave}
\DeclareAcronym{BAW}{short = BAW, long = bulk acoustic wave}
\DeclareAcronym{PA}{short = PA, long = power amplifiier}
\usepackage{cite}

\begin{document}
\title{Chirp-Based Multi-Device Ambient Backscatter Communication and Sensing Enabled by OFDM-AFDM Symbiotic Radio}
\author{Katia Abtouche, Fikiri Salum Uledi, Ayoub Ammar Boudjelal, Muhammad Bilal Janjua, Zinat Behdad, Cicek Cavdar,  and H\"{u}seyin Arslan, ~\IEEEmembership {Fellow,~IEEE}
        
\thanks{K.Abtouche, F. S Uledi,  A. A Boudjelal and H. Arslan are with the Department of Electrical and Electronics Engineering, Istanbul Medipol University, Istanbul, 34810, T\"{u}rkiye (email: abtouche.katia@std.medipol.edu.tr, fikiri.uledi@std.medipol.edu.tr, ayoub.ammar@std.medipol.edu.tr, huseyinarslan@medipol.edu.tr). 
\newline M. B. Janjua is with the R\&D Department, Oredata, Istanbul, T\"{u}rkiye, and also with T\"{u}rk Telekom R\&D Department, Istanbul, T\"{u}rkiye and (email: bilal.janjua@ieee.org). 
\newline Zinat Behdad and Cicek Cavdar are with the Department of Communication Systems, KTH Royal Institute of Technology, 100 44 Stockholm, Sweden (email: \{zinatb, cavdar\}@kth.se).}  
}
\maketitle
\begin{abstract}
This paper presents a novel symbiotic radio system for integrated sensing and backscatter communication (ISABC) technique that enables signal-domain interference-free coexistence of the primary communication signal and the \ac{BC} signal within the same spectrum. The proposed system design allows simultaneous \acp{BD} sensing and data transmission without mutual interference by exploiting waveform-domain orthogonality between \ac{OFDM} and \ac{AFDM} signals. Specifically, a chirp-based AFDM waveform is adopted due to its inherent processing gain, which enhances the detectability and reliability of the weak backscatter signal while simultaneously supporting high-resolution sensing. Unlike conventional methods that attempt to suppress direct-link interference (DLI), this approach embeds the backscatter transmission within the affine domain while maintaining reliable OFDM-based primary communication. Furthermore, by assigning distinct affine-domain shifts to each backscatter device, the proposed framework inherently suppresses \ac{IBDI}. Comprehensive simulation results demonstrate that the proposed coexistence scheme effectively mitigates interference without affecting the error rate of the primary link and improves the miss-detection probability performance of the \ac{BC}, making it a promising candidate for future low-power and interference-resilient systems.

\end{abstract}

\maketitle
\section{INTRODUCTION}
\label{sec:intro}
 
\IEEEPARstart{T}{he} rapid proliferation of the \ac{IoT} in modern wireless communication networks has led to an unprecedented increase in the number of connected devices, which is projected to exceed 40 billion by 2030 \cite{Ericsson_IoT_Connections_Outlook_2024}. This massive deployment introduces several critical challenges, including excessive energy consumption, spectrum congestion, and maintenance overhead, all of which hinder the scalability of \ac{IoT} networks. To address these challenges, the \ac{3GPP} through Release 19 has introduced three classes of \ac{A-IoT} devices categorized according to their energy storage capability, communication mechanism, design complexity, and power consumption. Device 1 (passive) and Device 2a (semi-passive) consume approximately 1 µW and are referred to as \acp{BD}, while Device 2b, consuming a few hundred µW, is classified as an active device \cite{3GPP_ref1,3GPP_PhY_Standardization_Overview}. Unlike conventional transmitters, \acp{BD} eliminate the need for power-intensive components such as amplifiers and oscillators \cite{van2018ambient}. Instead, they communicate by reflecting ambient \ac{RF} signals from surrounding sources, such as TV and AM/FM towers, rather than generating their own. The passive and semi-passive devices use a communication paradigm known as backscatter communication (\ac{BC}) \cite{AmBC_ContemSurv} which makes them more energy efficient than active devices.

However, the passive design and the absence of explicit coordination with the signal source makes \acp{BD} highly susceptible to signal fluctuations that degrade link quality. Another major performance bottleneck arises from \ac{DLI}, where the strong direct primary transmission interferes with the much weaker backscattered signal at the \ac{Rx} \cite{BDs_Standardization_Potentials_andChallenges} \cite{An_Overview_on_BC}. Moreover, in the massive deployment of multiple \acp{BD} for different data collection, an inter-backscatter-device interference (\ac{IBDI}) arises, where one \ac{BD} transmission causes interference to another \ac{BD} transmission \cite{tao2024integrated} \cite{long2019symbiotic}. The development of a low-power and reliable \ac{A-IoT} system requires to address these interference issues.

Given that contemporary cellular and Wi-Fi systems predominantly employ orthogonal frequency division multiplexing (\ac{OFDM}), ambient \ac{BC} can exploit the signal from the \ac{BS} to mitigate interference. Prior research has leveraged various \ac{OFDM}-domain techniques to address this challenge. For instance, in \cite{Subcarrier-wise_BC}, a subcarrier-wise \ac{OOK} scheme employing passive notch filtering was proposed, wherein a group of selected and filtered subcarriers were utilized for \ac{BC}. Although effective, this approach increases hardware and computational complexity and remains prone to leakage induced by \ac{ICI}. In \cite{Modulation_in_the_Air} \cite{ref18}, \ac{DLI} was mitigated by exploiting the uncorrupted portion of the \ac{CP} for backscatter signal transmission. While conceptually effective, the performance is highly sensitive to noise and power fluctuations. Although the authors extended their design to multi-antenna operation, it requires computationally heavy optimization for antenna combining. Similarly, the pilot-aided \ac{DSK} approach in \cite{pilot_aided_DSK} leveraged pilot signals for modulation and optimal detection. However, pilot reuse can compromise synchronization and channel estimation performance. In \cite{yang2018cooperative}, the cooperative \ac{Rx} was designed to eliminate DLI by detecting the RF source signal first and removing it through \ac{SIC}. However, this system depends on good cooperation between the RF source and the BD and also requires very accurate detection in the first SIC stage, making it sensitive to error propagation. Despite all research advancements, ambient \ac{BC} remains inherently interference-limited due to the lack of coordination and control over the primary signal source \cite{AmBC_ContemSurv}. To overcome this limitation, the \ac{SBC} framework was proposed in \cite{Janjua_SymbioticRad,SRad} \cite{uledi2025interference}. In this paradigm, the primary signal design and \ac{BS} operations not only serve the primary users but also support the \acp{BD} within the coverage area in a coordinated manner, transforming \acp{BD} from passive listeners into cooperative network elements. In \ac{SBC} methodologies, the \ac{BS} adapts its primary signal design to account for the presence of \acp{BD}, thereby mitigating \ac{DLI} and \ac{IBDI} and ultimately improving overall system detection performance. A representative example is the scheme in \cite{janjua2025interference}, where the BS deliberately allocates a dedicated empty subcarriers that BDs can shift their reflections onto. By confining all backscatter-induced frequency shifts to this reserved subcarrier, the receiver can easily discriminate BD signals from the primary OFDM data, achieving clean separation in the frequency domain.

To further enable context-aware \ac{IoT} applications such as smart logistics, healthcare, smart homes, and other emerging \ac{IoT} use cases, sixth-generation (6G) design requires both energy-efficient communication and integrated sensing functionalities. Consequently, \ac{ISABC} has emerged as a promising paradigm that builds upon \ac{BC}’s ultra-low-power operation to achieve sensing capabilities with minimal power \cite{toro2021backscatte} \cite{kaushik2024integratedIoT_Sensing}. Several recent studies have investigated various aspects of \ac{ISABC}. In \cite{BISAC}, the authors proposed a typical \ac{ISABC} framework that demonstrates the potential of \acp{BD} to enhance conventional \ac{ISAC} performance in \ac{IoT} applications. Their study highlights how passive devices can effectively contribute to environmental sensing while maintaining low-power operation. The work in \cite{zargari2025transmit} explored transmit power optimization in \ac{ISAC} systems incorporating \acp{BD}, showing that appropriate power allocation can balance \ac{BC} and sensing performance while mitigating potential interference. Similarly, \cite{wang2023integrated} extended this concept to \ac{RIS}-assisted backscatter systems, illustrating how \ac{RIS} can simultaneously enhance sensing accuracy and communication throughput in passive networks. Furthermore, in \cite{tao2024integrated}, the authors proposed a \ac{SR} framework in which multiple \ac{IoT} devices backscatter the primary signal, jointly enabling data transmission and environmental sensing through coordinated multi-device access and signal processing. The authors in \cite{xu2024joint} propose a joint localization and signal detection framework for ambient OFDM-backscatter systems, where the receiver first estimates BD delays/angle-of-arrivals, then detects differentially encoded BD symbols. Although these studies provide valuable insights, they do not consider the impact of \ac{DLI} and \ac{IBDI} under limited \ac{RF} resources and rely on ideal interference cancellation, which remains a key challenge in achieving reliable \ac{BC} and accurate sensing.

Since the primary communication, \ac{BC}, and sensing share the same \ac{RF} resources, waveform redesign is essential to ensure efficient coexistence. The optimal approach lies in adopting a unified coexistence design capable of supporting both primary communication, \ac{BC}, and sensing within a single physical layer. In this context, the legacy of \ac{OFDM} for communication, together with the advantageous features of \ac{AFDM} and their mutual orthogonality, makes an \ac{OFDM}-\ac{AFDM}-based framework a promising candidate for achieving \ac{ISABC}-based symbiotique radio. 

By leveraging the chirp diversity and affine-domain properties of \ac{AFDM} within an OFDM coexistence framework, we propose a design that achieves enhanced robustness, reduced \ac{DLI}, and significantly improved detection of weak BD signals for large-scale \ac{ISABC} operation. While OFDM-AFDM coexistence has been studied previously as a joint waveform design \cite{BIR_2025}, this work is the first to exploit the inherent \ac{AFDM} structure specifically for backscatter delay-domain sensing and detection inside an OFDM system. The main contributions are summarized as follows:

\begin{itemize}
\item We demonstrate that \ac{AFDM}’s chirp-domain diversity, affine-domain sparsity, and constant-envelope structure can be harnessed to perform robust BD detection within an \ac{OFDM} system. Unlike prior coexistence studies that focus on communication performance, our design exploits \ac{AFDM}’s waveform characteristics to embed backscatter communication and sensing without violating OFDM orthogonality or generating cross-waveform interference. The chirp based structure in AFDM provides high delay resolution, strong post-compression energy focusing, and a constant-envelope profile well suited for extracting weak \ac{BD} signal.
\item We introduce a delay-shifting \ac{DSK}-\ac{OOK} backscatter strategy, in which each \ac{BD} applies a deterministic delay that maps its reflection to a clean region of the \ac{CP} beyond the primary system’s maximum excess delay. This simple waveform-driven mechanism enables many \ac{BD} to coexist without codebooks, spreading sequences, or multi-antenna processing, while inherently mitigating \ac{DLI} and \ac{IBDI} through affine-domain separability.

\item We demonstrate that \ac{AFDM}'s quadratic-phase chirp structure provides substantial processing gain for detecting low-power \ac{BD} reflections. After de-chirping, the backscattered energy collapses into a sharp, non-fading affine-domain peak, offering much greater stability and contrast than OFDM-only \ac{BC}. This property enables reliable non-coherent detection even under severe power asymmetry with the direct link, a regime where classical ambient OFDM backscatter techniques struggle.

\item We employ \ac{AFDM} pilot waveforms to estimate the round-trip delays of individual \acp{BD}, enabling precise range sensing and intentional affine-domain separation among devices. This differs from conventional \ac{AFDM}-based sensing, which typically characterizes only environmental paths; here, \ac{AFDM} is used to extract the \acp{BD}’ own delay signatures. We further develop a non-coherent detector and provide an extensive performance evaluation including probability of miss-detection (PMD), bit error rate (BER), sum-rate, and \ac{RMSE}, along with a one-shot efficiency criterion for tuning \ac{ISABC} parameters.

\item Finally, extensive numerical results demonstrate that the proposed \ac{ISABC} transceiver achieves a significantly lower miss-detection probability, improved sensing accuracy, and higher sum-rate compared to OFDM-only backscatter approaches. The simulations further confirm the practical effectiveness of the delay-domain \ac{AFDM} pilot for range estimation and show that the system maintains reliable performance across multiple \acp{BD}, signal-to-noise ratio (SNR) conditions, and different subcarrier configurations, verifying its suitability for real-world multi-device \ac{ISABC} symbiotic radio deployments.

\end{itemize}

The rest of the paper is organized as follows. Section \ref{sec:syst_mod} presents the design of the \ac{SR} system along with the foundational concepts applied throughout the subsequent sections. Section \ref{sec:proposed_scheme} describes the proposed \ac{ISABC} schemes. Section \ref{sec:pefromance_analysis} shows derivation of detection procedures and performance analysis. Section \ref{sec:sim_results}, presents the simulation results and discusses the theoretical findings. Conclusions and future works are provided in Section \ref{sec:concl}.

{\em Notation:} $\mathbb{Z}^+$ represents the set of positive integers, $\mathcal{CN}(0, \sigma^2)$  denotes the circularly symmetric complex Gaussian distribution with zero mean and variance $\sigma^2$, while $\mathbb{E}[\cdot]$ represents the expectation of its argument over random variables. The probability of an event $A$ is denoted by $\Pr(A)$, while the conditional probability of $A$ given $B$ is expressed as $\Pr(A \mid B)$. The notation $X \mid \mathcal{H}_i$ denotes the random variable $X$ 
conditioned on hypothesis $\mathcal{H}_i$. The
$\chi^{2}_{L}$ denotes a central chi-square random variable with $L$ degrees of freedom, while $\chi^{2}_{L}(\lambda)$ denotes a non-central chi-square random variable with 
$L$ degrees of freedom and non-centrality parameter $\lambda$.
For a constant $c>0$, the notation $c\,\chi^{2}_{d}$ denotes a 
scaled chi-square random variable, i.e., if $X\sim\chi^{2}_{L}$ 
then $c\,\chi^{2}_{L}$ represents the random variable $cX$.

\section{SYSTEM MODEL}
\label{sec:syst_mod}
\begin{figure}[t]
    \centering
\includegraphics[width=0.5\textwidth]{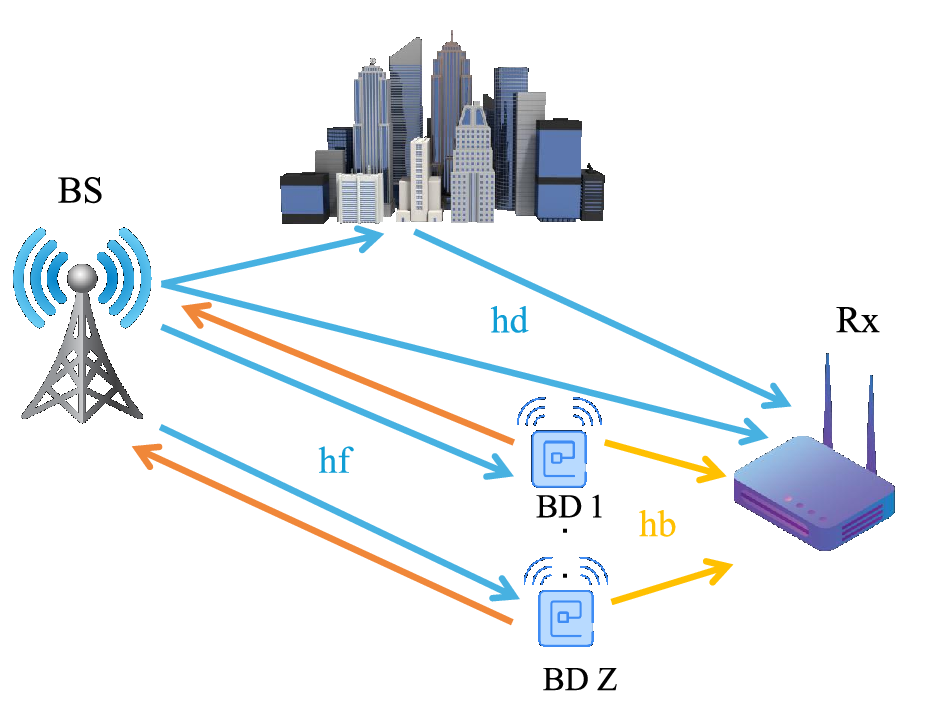}
  \caption{The proposed ISAC OFDM-AFDM based symbiotic radio system model.}
    \label{fig:sysmod}
\end{figure}

Consider an \ac{SR} system in which the \ac{BS} transmits a unified \ac{OFDM}-\ac{AFDM} signal that illuminates both the \ac{Rx} and a set of Z passive \acp{BD}. These \ac{BD}s, categorized as Type-Device 1 \ac{A-IoT} nodes, modulate and reflect the incident signal toward the \ac{Rx}, thereby enabling ultra-low-power communication. In this configuration, the \ac{BS} employs the AFDM pilot structure to perform monostatic environmental sensing from the reflected components. Consequently, the \ac{Rx} thus receives a superposition of the direct OFDM link and the BD-reflected signals, as depicted in Fig.\ref{fig:sysmod}.
\subsection{Primary Signal}
The \ac{BS} transmits a unified \ac{OFDM}-\ac{AFDM} signal in which the \ac{OFDM} signal is composed of $N$ subcarriers. The corresponding time-domain signal generated using an \ac{IDFT} is given as
\cite{hwang2008ofdm}
\begin{equation}
   \begin{aligned}
s^{\text{OFDM}}[n] = \frac{1}{\sqrt{N}} \sum_{m=0}^{N-1} X[m] e^{j \frac{2\pi mn}{N}}, 
\label{eqn:prim_ofm}
\end{aligned} 
\end{equation}
here, $m$ denotes the frequency-domain subcarrier index and $n$ denotes the corresponding time-domain sample index,
and $ N $ is the \ac{IDFT} size. A \ac{CP}, longer than the channel delay spread is appended to mitigate \ac{ISI}.

An \ac{AFDM} symbol is defined as a one-dimensional vector of data symbols $X \in \mathbb{C}^{N \times 1}$, which is directly multiplexed into a twisted time-frequency affine domain using the \ac{IDAFT}. The \ac{AFDM} part enables \ac{BC} with the same user \ac{Rx}. In discrete form, an \ac{AFDM} symbol can be written as \cite{bemani2023affine}

\begin{equation}
   \begin{aligned}
    P_{time}[n] =  \frac{1}{\sqrt{N}} X^{\text{AFDM}}[i] e^{\jmath 2 \pi\left(c_1 n^2+ni /N+ c_2 i^2\right)},
    \label{equ: AFDM}
\end{aligned} 
\end{equation}
where $i$ denotes the active affine-domain index, and $c_{1},\, c_{2}$ are the \ac{AFDM} parameters selected for delay-domain resilience. The final composite unified signal transmitted by the \ac{BS} is expressed as
\begin{equation}
\label{eqn:comb}
    s_p[n] = s^{\text{OFDM}}[n] + P_{time}[n].
\end{equation}

\subsection{\ac{BD} Signal}
The \ac{BD} architecture exploits the intrinsic \ac{RF}-acoustic transduction properties of \ac{SAW}/\ac{BAW} delay-line circuits. A reflection-control switch selects one of the $M$ available delay paths as a function of the current bit or symbol \cite{ref30}. Because acoustic wavelengths are much shorter than their electromagnetic counterparts, these circuits substantially reduce propagation velocity in the acoustic domain and then reconvert the signal to the electrical domain, thereby realizing delays on the order of hundreds of nanoseconds within compact, sub-centimeter footprints \cite{ref19}. Under \ac{DSK}, a transmitted bit `0' corresponds to an absorptive impedance with negligible \ac{BD} signal, whereas a bit `1' engages a deterministic delay and reflects a controlled portion of the incident waveform toward the \ac{Rx}. The received signal at ${z}^{\text{th}}$ \ac{BD} propagates through a frequency-selective channel with $L_f$ taps. Its discrete-time representation is
\begin{equation}
y_{\mathrm{b,z}}[n]
=
\sum_{f=0}^{L_f-1}
h_f\; s_p[n-\ell_f],
\end{equation}
where $h_f$ and $\ell_f$ denote the gain and delay of the $f$th tap of
the \ac{BS} to {BD} channel, respectively.

The BD then imposes its modulation through a delay operation characterized by
$\ell_{BD}$ and re-radiates a portion of the impinging signal. The resulting
backscattered signal is expressed as
\begin{equation}
x_{z}[n]
=
\alpha\, b[\kappa] y_{b,z}[n-\ell_{BD,z}],
\end{equation}
with $\alpha$ being the reflection coefficient and $y_b[\cdot]$ the delay-based signature of the \ac{BD}, while $b[\kappa]\in\{0,1\}$ represents the \ac{BD}'s information data.

\begin{figure*}[t]
    \centering    \includegraphics[width=0.8\textwidth]{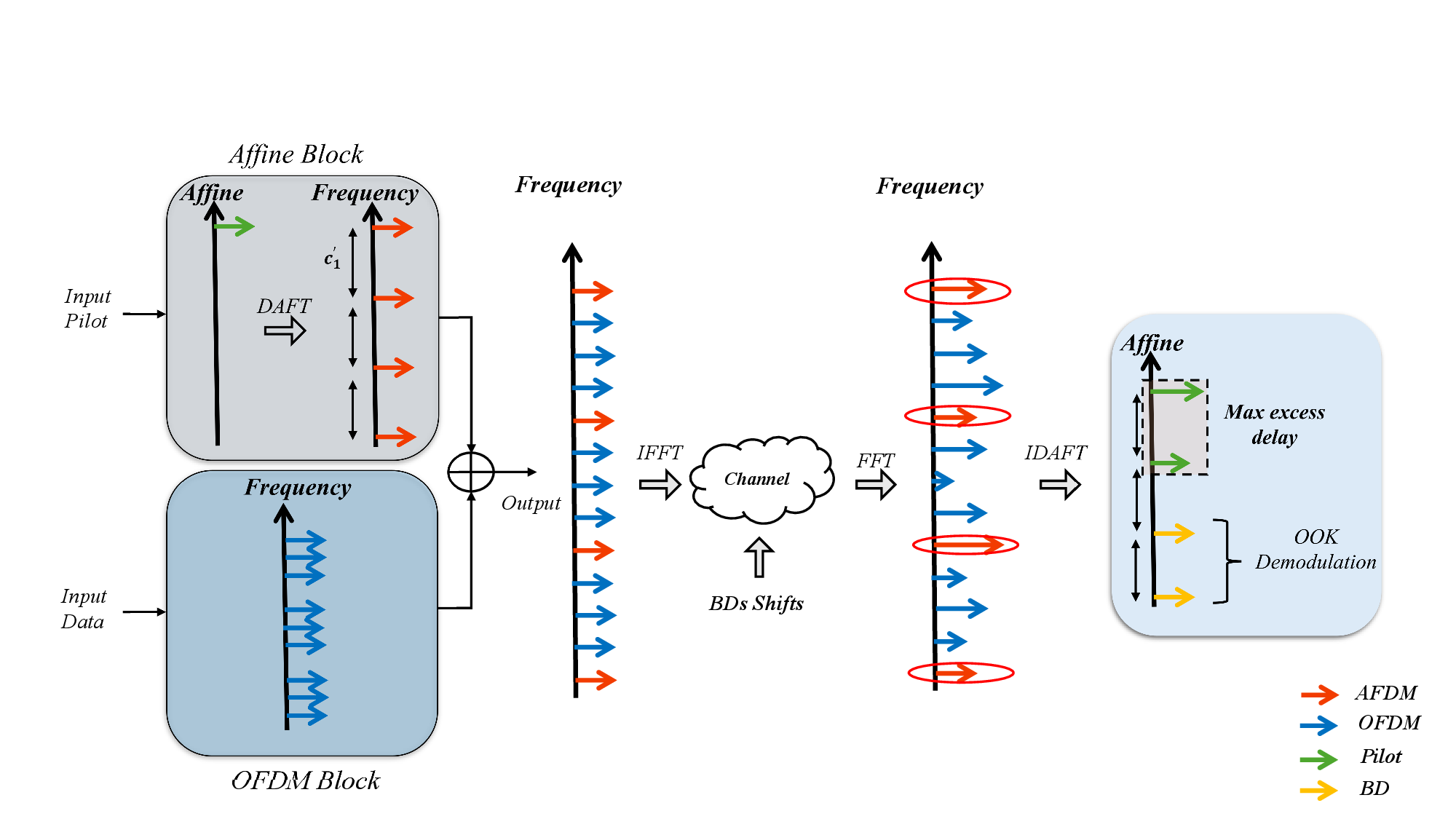}
    \caption{Illustration of the proposed pilot design.}
    \label{fig:pilot_des}
\end{figure*} 
\subsection{Received signal}
At the \ac{Rx}, the received signal consists of the direct link signal and the backscatter signal, which is expressed as
\begin{equation}
    y_p[n] = h_d[n] \ast s_p[n] + \sum_{z=1}^{Z} h_{b,z}[n] \ast x_z[n] + w[n],
    \label{eqn:BD_mod}
\end{equation}
where $w \sim CN(0, \sigma^2)$ represents \ac{AWGN}, $h_d[n]$ is the direct link channel and $h_{b,z}[n]$ is the backscatter channel, which models the single-path propagation from the BD to the BS. The $h_d$ is modeled as multipath of the reflection by the objects in the environment other than the \acp{BD} and the direct link only. This channel is regarded as a linear time invariant (LTI) channel expressed as 
\begin{equation}
   \begin{aligned}
h_{d}[n] = \sum_{d=0}^{D-1} h_d \delta[n - \ell_{d}],
\end{aligned} 
\end{equation}
where $h_d$ and $\ell_d$ denotes the ${d}^{\text{th}}$ tap gain and delay value of the BS to RX channel consisting of D taps, respectively. 

\section{PROPOSED \ac{ISABC} SCHEME}
\label{sec:proposed_scheme}
The proposed \ac{ISABC} framework jointly leverages the high spectral efficiency and broadband data delivery capability of \ac{OFDM}, along with the resilience to doubly selective channels and sensing suitability of \ac{AFDM}, within a unified physical-layer design. In the proposed architecture, the primary communication is conveyed using conventional \ac{OFDM} symbols confined to the frequency-domain, whereas the ambient \ac{BC} and sensing signals are embedded in the affine domain via \ac{AFDM}. This domain separation renders \ac{BD} transmissions inherently distinguishable from the primary \ac{OFDM} signal, without incurring additional spectral overhead or invoking heavyweight interference-cancellation mechanisms. The unified framework and signal processing is modeled through several steps as illustrated in Fig.~\ref{fig:pilot_des}.

\subsection{Pilot and data allocation}

The proposed framework introduces a dual-purpose pilot designed in the affine domain. In this design, the pilot is reflected by the \acp{BD} toward the \ac{Rx} to enable \ac{BC}, while the same pilot is retro-reflected to the \ac{BS} in a monostatic radar configuration to facilitate range estimation. In parallel, the same pilot in the frequency-domain is leveraged for \ac{CSI} acquisition for the primary communication link. The pilot’s representation in these two domains is given as:
\subsubsection{Pilot signal in time-domain representation}

\begin{figure}[t]  
    \centering
   \includegraphics[width=0.7\linewidth]{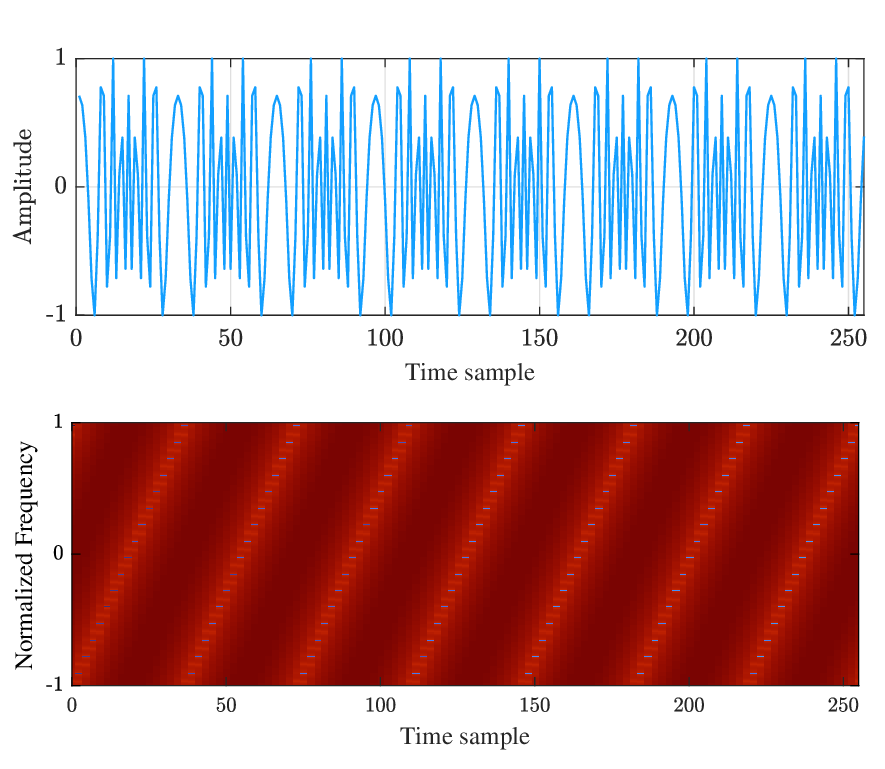} 
    \caption{Time-domain AFDM pilot signal and its corresponding time-frequency representation with $c_1'$=8.}
    \label{fig:figure3}
\end{figure}

The time-domain AFDM signal is generated using the \ac{IDAFT} as 
\begin{equation}
P_{time}[n] =
\begin{cases} 
\sqrt{P_{pilot}} X^{\text{AFDM}}[i] e^{\jmath 2 \pi\left(c_1 n^2+ni /N+ c_2 i^2\right)}, &  i = 1. \\
0,  \text{otherwise}.
\end{cases} 
\label{equ:affine_comm}
\end{equation}
where $P_{pilot}$ is the pilot power, the term \(e^{j2\pi c_{1}n^{2}}\) imposes a quadratic phase profile; in discrete-time signals, a quadratic phase corresponds to a chirp \cite{bemani2024integrated}. Hence, AFDM inherently produces a chirp-like waveform irrespective of the data symbols as shown in \figurename~\ref{fig:figure3}.
To strengthen the chirp structure, the AFDM design parameter is constrained as
\begin{equation}
c_{1} = \frac{c_{1}'}{2N}, \qquad c_{1}' \in 2\mathbb{Z}^{+},
\label{eq:c1_constraint}
\end{equation}
which allows expressing the AFDM block size as \(N=c_{1}'M\), where $c_{1}'$ and $M$  being a power of 2. Under this condition, the quadratic phase cycles every \(M\) samples, making the AFDM waveform equivalent to \(c_{1}'\) repetitions of the same discrete chirp of length \(M\) as given in \cite{BIR_2025} by
\begin{equation}
s_{\mathrm{chirp}}[\mu
]=e^{j\pi\frac{\mu
^{2}}{M}}, \qquad \mu
=0,1,\dots,M-1.
\label{eq:chirp}
\end{equation}
where $\mu$ is simply the time sample index within one chirp segment. 

\subsubsection{Pilot signal in frequency-domain representation}
The frequency-domain AFDM pilot is obtained by applying the \ac{DFT} as
\begin{equation}
\begin{aligned}
    P_{freq}[m] = & \frac{1}{\sqrt{N}} \sum_{n=0}^{N-1}{P_{time}[n]} e^{-j 2 \pi \frac{n m}{N}}.
    \label{equ:frq_dom}
\end{aligned}
\end{equation}
substituting~(\ref{equ:affine_comm}) into ~(\ref{equ:frq_dom}), this representation becomes
\begin{equation}
\begin{aligned}
P_{\mathrm{freq}}[m]
&= \frac{\sqrt{P_{\mathrm{pilot}}}}{\sqrt{N}}
X^{\mathrm{AFDM}}(i)\, e^{j 2\pi c_2 i^2} \\
&\quad \times
\sum_{n=0}^{N-1}
e^{j 2\pi \left( c_1 n^2 + \frac{n(i-m)}{N} \right)}
\end{aligned}
\label{equ:frq_exp}
\end{equation}
substituting (\ref{eq:c1_constraint}) into (\ref{equ:frq_exp}), this representation becomes
\begin{equation}
\begin{aligned}
P_{\mathrm{freq}}[m]
&= \frac{1}{\sqrt{N}} \sqrt{P_{\mathrm{pilot}}}\, X^{\mathrm{AFDM}}(i)\,
e^{j 2\pi c_2 i^2} \\
&\quad \times
\left(
\sum_{n=0}^{N-1}
e^{j \pi \frac{c_1'}{N} n^2}
e^{j \frac{2\pi}{N} n i}
e^{-j \frac{2\pi}{N} n m}
\right)
\label{equ:fin_frq_exp}
\end{aligned}
\end{equation}
The inner sum in~(\ref{equ:fin_frq_exp}) is the \ac{DFT} of a chirp sequence that is periodically repeated $c_1'$ times across the $N$ samples. By standard \ac{DFT} identities, periodic repetition in the time-domain produces an $M$-point upsampling in the frequency-domain, which concentrates the spectral energy on every ${c_1'}^{\text{th}}$
subcarrier. Consequently, the subcarriers bearing the \ac{AFDM} pilot form a sparse combination in the frequency-domain, with non-zero tones appearing every $c_1'$ subcarrier, leaving the interleaved tones unused by \ac{AFDM}. Conversely, the \ac{OFDM} waveform places data exclusively on these complementary frequency bins, with its frequency-domain vector defined as the inverse of~(\ref{eqn:prim_ofm}). This disjoint allocation provides a unified waveform framework that ensures strict orthogonality between \ac{AFDM} and \ac{OFDM}, as explained in \cite{BIR_2025}
\begin{equation}
 \sum_{m=0}^{N-1} P_{freq}[m] X^{\text{OFDM}}[m] = 0~.
\end{equation}

\subsection{\ac{BD} modulation}
Each \ac{BD} conveys information by deterministic \ac{DSK} applied to the incident unified \ac{OFDM}-\ac{AFDM} waveform. Consider $b[\kappa]\in\{0,1\}$ denoting the $\kappa^\text{th}$ bit of the $j^\text{th}$ \ac{BD}. For one \ac{OFDM}-\ac{AFDM} block in $s_p[n]$, the baseband signal produced by \ac{BD} is given as
\begin{equation}
x_{z}[n] \;=\;
\begin{cases}
0, & b[\kappa]=0, \\[2pt]
\ \alpha\,
        s_p[n - \ell_{fmax}-\ell_{\mathrm{BD},z}], & b[\kappa]=1.
\end{cases}
\label{eq:bd_baseband2}
\end{equation}
At the \ac{Rx}, the aggragated sigal from BDs is given   
\begin{equation}
s_{BD}[n] = \sum_{z=1}^{Z} x_{z}[n] * h_{b,z}[n].
\label{eq:bd_rx}
\end{equation}
To ensure that \ac{BC} does not contaminate the primary \ac{OFDM} data, the \ac{BD} delay is confined to the clean \ac{CP} region, which is 
\begin{equation}
\ell_{d\max} \;<\; \ell_{BD,Z} \;<\; CP \ length.
\label{eq:cp_constraint}
\end{equation}
Therefore, for \(J\) simultaneous BDs indexed by \(j\), we assign pairwise distinct
delays \(\ell_{BD,z}\) such that
\begin{equation}
\ell_{BD,Z} - \ell_{BD,Z-1} \;\ge\; \Delta\ell_{\min},
\tag{16}
\end{equation}
where the effective minimum delay separation is
\begin{equation}
\Delta\ell_{\min} = \Delta\tau + L_f + 1.
\end{equation}
Here, \(\Delta\tau \ge 1\) ensures one-sample resolvability and is
practically chosen larger than the dominant spread of each BD path
to enhance separability.
Under the delay spacing rule in (16), the maximum number of BDs
that can be accommodated is given by

\begin{equation}
Z_{\max} \;\approx\;
\min\!\left(
\left\lfloor
\frac{CP - \ell_{d\max}}{\Delta \ell_{\min}}
\right\rfloor,
\;
\left\lfloor
\frac{(\dfrac{N}{c_1'} - D )+ 1)}{L_f + 1}
\right\rfloor
\right).
\label{equ:max_bd}
\end{equation}
The equation~(\ref{equ:max_bd}) explicitly shows how the CP length, the AFDM spreading factor~$c_1'$, the number of
subcarriers~$N$, and the channel spread jointly determine the
number of coexisting BDs supported by the system.

\begin{figure*}[t]  
    \centering   \includegraphics[width=0.6\linewidth]{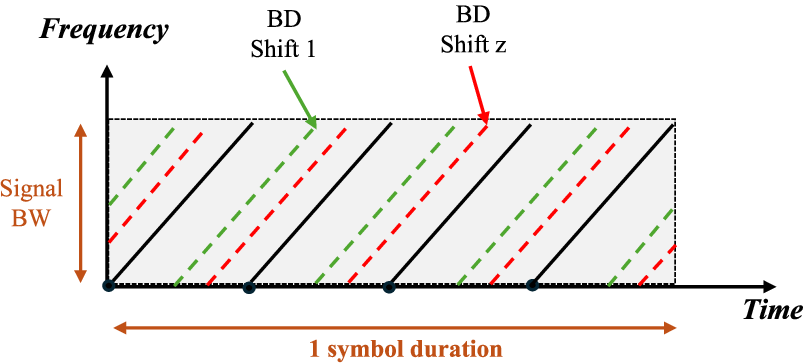} 
    \caption{The time-frequency illustration of the incident AFDM signal at the BD with DSK modulation performed by the BDs  with $c_1'$=4.}
    \label{fig:BD_chirp_shifts}
\end{figure*}

The AFDM component of $s_p[n]$ consists of identical chirp-like patterns, whose time-shifted versions preserve the same structure. Fig.~\ref{fig:BD_chirp_shifts} illustrates how the AFDM pilot, composed of repeated chirps, undergoes distinct delay shifts at different BDs,
resulting in shifted chirp replicas at the BD output. Consequently, each delay $\ell_{BD,z}$, introduced by a BD, preserves the AFDM waveform and maps to a sparse, non-overlapping affine-domain shift of size $c_1'\ell_{BD,z}$ on the AFDM pilot. The reflection coefficient~$\alpha$ scales the BD
link strength linearly in amplitude and directly affects the received \ac{SNR}. The use of chirp-based AFDM pilots is particularly advantageous for delay-shift keying in backscatter links. Chirps preserve their energy under integer-sample time shifts, meaning that even a very weak BD reflection produces a distinct and concentrated signature after the AFDM transform. This property significantly improves the reliability of non-coherent OOK detection at the \ac{Rx}, especially in low-SNR conditions where conventional tone-based pilots suffer from severe energy spreading. Because the energy of a chirp is distributed in time but recompressed into a sharp affine-domain peak, the BD-induced delay shift remains detectable even after round-trip attenuation and the additional BD reflection loss. These characteristics make chirp-driven
DSK inherently robust to fading and power limitations typical of passive backscatter devices. In practice, $\alpha,\,\ell_{BD,z}$ and the
AFDM pilot power are jointly configured to achieve a target miss-detection probability under a specified false-alarm constraint.

\section{Detection and Sensing Performance Analysis}
\label{sec:pefromance_analysis}

At the \ac{Rx}, received baseband signal consists of the direct-link component and the superposition of multi-\acp{BD} backscatter reflections, represented as
\begin{equation}
y[n]=
\sum_{d=0}^{D-1} h_d s_p[n-\ell_d]
+
 s_{BD}[n]
+
\omega[n]~,
\label{eq:rx_time}
\end{equation}
All delays are assumed integer-valued so that they correspond to exact affine-domain shifts under the AFDM transform. In \eqref{eq:rx_time}, $s_p[n-\ell_d]$ is dedicated to primary communication, whereas the $s_{BD}[n]$, leveraging its chirp-like behavior, is processed digitally for multi-\acp{BD} data detection at the \ac{Rx} and for sensing at the \ac{BS}. For the primary link, the \ac{Rx} performs conventional coherent \ac{OFDM} detection, with direct-link \ac{CSI} obtained via pilot-aided channel estimation as in standard systems. However, in this case, the pilots are defined in the affine domain. In contrast, the multi-\acp{BD} symbols are recovered using low complexity and low computational non-coherent detection approach in the affine domain which is orthogonal from primary signal which is in \ac{OFDM} domain  \cite{NonCohrent_Darsena} \cite{Noncoherent_Qian}.
\subsection{Non-coherent detection for the multi-\acp{BD} signal}\
The \ac{AFDM} waveform naturally exhibits a sparse structure due to the \ac{AFDM} modulation parameters; non-zero affine-domain coefficients appear only at indices spaced by the factor $c_1'$, leaving most bins without AFDM contributions. This sparsity is crucial for detection, since each \ac{BD} introduces a deterministic propagation delay into the incoming $s_p[n-\ell_f]$, resulting in a predictable affine-domain shift. In this regard, each time-domain sample of $s_{BD}[n]$ is related to its affine-domain representation through the \ac{AFDM} transform at index $i$ and provides
\begin{equation}
Y^{\text{AFDM}}[i]
=
\frac{1}{N}
\sum_{n=0}^{N-1}
y[n]\,
e^{-j\frac{2\pi}{N}\left(c_1 n^2 + n i + c_2 i^2\right)},
\end{equation}
therefore,
\begin{equation}
\begin{aligned}
Y^{\mathrm{AFDM}}[i]
&=
\frac{1}{N}
\sum_{i'=0}^{N-1}
X^{\mathrm{AFDM}}[i']  \\
&\quad \times
e^{j\frac{2\pi}{N}
\left(
c_2(i'^2-i^2)+c_1\ell^2
+(i'-i-c_1'\ell)n-i'\ell
\right)}
\end{aligned}
\label{eq:inverse_AFDM}
\end{equation}
where $i'$ is the input affine-domain index, $i$ is the output index after transformation, and variable $\ell$ denotes a total delay, which is later replaced by $\ell_d$ and $\ell_{BD_{total},z}$ that represente the total \ac{BD} delay in \eqref{eq: ed1} and \eqref{eq:ed2}, respectively. Equation~\eqref{eq:inverse_AFDM} can be rewritten as 

\begin{equation}
\begin{aligned}
Y^{\mathrm{AFDM}}[i]
&= \frac{1}{N} \sum_{n=0}^{N-1} \Bigg[
\sum_{d=0}^{D-1}\sum_{i'=0}^{N-1}
h_d X^{\mathrm{AFDM}}[i'] e^{j\Xi_d} \\
&\quad + \sum_{z=1}^{Z} \sum_{f=0}^{L_f-1} \sum_{i'=0}^{N-1}
\alpha\, h_{f,z} h_{b,z} X^{\mathrm{AFDM}}[i'] e^{j\Xi_{\mathrm{BD},z}}
\Bigg] \\
&\quad + \Omega(i)
\end{aligned}
\label{eq:AFDM_sum}
\end{equation}
where $\Omega(i)$ is AWGN in the affine domain, with
\begin{equation}
\Xi_d
=
\frac{2\pi}{N}
\left[
c_1 \ell_d^2
+
c_2 \left(i'^2 - i^2\right)
+
\left(i' - i - c_1' \ell_d\right) n
-
i' \ell_d
\right],
\label{eq: ed1}
\end{equation}
\begin{equation}
\begin{aligned}
\Xi_{\mathrm{BD},z}
&= \frac{2\pi}{N}
\Big[
c_1 \ell_{\mathrm{BD}_{\mathrm{total},z}}^2
+ c_2 (i'^2 - i^2) \\
&\quad + (i' - i - c_1' \ell_{\mathrm{BD}_{\mathrm{total},z}}) n
- i' \ell_{\mathrm{BD}_{\mathrm{total},z}}
\Big]
\end{aligned}
\label{eq:ed2}
\end{equation}

The exponential term $e^{j\frac{2\pi}{N}(i'-i-c_1'\ell)n}$, depending on $n$, defines the finite geometric sum as 
\begin{equation}
S_N(i',i,\ell)
=
\sum_{n=0}^{N-1}
e^{j\frac{2\pi}{N}(i'-i-c_1'\ell)n},
\label{eq:geom_sum}
\end{equation}
which is equivalent to $S_N(i',i,\ell)=
N\,\delta_N(i'-i-c_1'\ell)$, this simply states that the expression is non-zero only when the affine-domain index $i'$ matches the shifted index $i + c_1' \ell$ modulo $N$. This follows from the role of $\delta_N(\cdot)$, the periodic kronecker delta, which evaluates to one only when its argument wraps around to zero under modulo-$N$ arithmetic and becomes zero otherwise. As a result, only those indices satisfying $i' \equiv i + c_1' \ell \pmod{N}$ contribute to the summation, while all other terms vanish. To simplify the notation and capture these shifts more cleanly, we define the aligned affine-domain indices $i_d = i + c_1' \ell_d$ for the direct path and $i_{\mathrm{BD},z} = i + c_1' \ell_{BD_{total},z}$ for the $z$-th backscatter device. With these substitutions, \eqref{eq:AFDM_sum} reduces to 

\begin{equation}
\begin{aligned}
Y^{\text{AFDM}}[i]
&=
\sum_{d=0}^{D-1}
h_d\, X^{\text{AFDM}}[i_d]e^{j\Phi_d[i]}
\\
&\quad
+
\sum_{z=0}^{Z-1} \alpha\ \sum_{f=0}^{Lf-1}
h_{f,z} \ h_{b,z}\, X^{\text{AFDM}}[i_{BD,z}]e^{j\Phi_{BD,z}[i]}  
\\
&\quad
+\,\Omega[i],
\end{aligned}
\label{eq:Y_affine_u}
\end{equation}
where
\begin{equation}
\Phi_d[i] 
= \frac{2\pi}{N}
\Big[
c_1 \ell_d^2 
+ c_2\big((i+c_1'\ell_d)^2 - i^2\big)
- (i + c_1'\ell_d)\,\ell_d
\Big],
\label{eq:phi_d}
\end{equation}
\begin{equation}
\Phi_{\mathrm{BD},z}[i]
=
\begin{aligned}[t]
&\frac{2\pi}{N}\Big[
c_1 \ell_{\mathrm{BD}_{\mathrm{total},z}}^2
+ c_2\big((i+c_1'\ell_{\mathrm{BD}_{\mathrm{total},z}})^2-i^2\big)
 \\
&\quad
- (i+c_1'\ell_{\mathrm{BD}_{\mathrm{total},z}})\,\ell_{\mathrm{BD}_{\mathrm{total},z}}
\Big]
\end{aligned}
\label{eq:phi_bd}
\end{equation}
We therefore obtain the affine-domain input-output relation as
\begin{equation}
\begin{aligned}
Y^{\mathrm{AFDM}}[i]
&= \sum_{d=0}^{D-1}
h_d\, X^{\mathrm{AFDM}}(i+c_1'\ell_d)\,
e^{j\Phi_d[i]} \\
&\quad + \sum_{f=0}^{L_f-1} \sum_{z=0}^{Z-1}
\alpha\, h_{f,z} h_{b,z} \\
&\qquad \times
X^{\mathrm{AFDM}}(i+c_1'\ell_{\mathrm{BD}_{\mathrm{total},z}})
\, e^{j\Phi_{\mathrm{BD},z}[i]} \\
&\quad + \Omega[i]
\end{aligned}
\label{eq:Y_affine_i}
\end{equation}
showing that each BD contributes a shifted cluster of non-zero affine-domain coefficients.

Let $\mathcal{I}_{K,z} = \{ k = 1, 2, \ldots, K \}$ denote the set of affine-domain indices occupied by the $z$-th BD, where the index location is determined by the intentional affine-domain shift, i.e.,$ k = i + c_1' \ell_{\mathrm{BD,total},z}$.
Detection is performed independently for each BD. To establish the hypothesis testing framework, a single representative index from the corresponding affine-domain cluster is considered. Without loss of generality, the first index of the set $\mathcal{I}_{K,z}$ is selected to define the observation model.

The affine-domain observation for BD $z$ at index $k \in \mathcal{I}_{K,z}$ is denoted by $Y_{\mathrm{BD}}^{\mathrm{Affine}}[k]$. Since the hypotheses differ only by the presence or absence of the deterministic BD component \cite{guruacharya2020optimal}, the detection decision model is primarily expressed as

\begin{equation}
Y^{\mathrm{Affine}}_{\mathrm{BD}}[k] =
\begin{cases}
W[k], & \mathcal{H}_0,\\[3pt]
S[k] + W[k], & \mathcal{H}_1,
\end{cases}
\qquad k\in I_{K,z},
\end{equation}
where $S[k]$ is the deterministic \ac{BD}$_z$ contribution and $W[k]\sim\mathcal{CN}(0,\sigma^2)$.

The corresponding energy-based non-coherent detector forms the decision statistic expressed as
\begin{equation}
E_z = \sum_{k \in I_{K,z}} \big| Y^{\mathrm{Affine}}_{\mathrm{BD}}[k] \big|^2.
\end{equation}

Under $\mathcal{H}_0$ hypothesis, \ac{BD}$_k$ is inactive and $Y^{\mathrm{Affine}}_{\mathrm{BD}}[k] = W[k]$ 
Hence, each $|Y^{\mathrm{Affine}}_{\mathrm{BD}}[k]|^2$ is exponentially distributed, and the sum $E_k$ follows a central chi-square law with $2K$ degrees of freedom
\begin{equation}
E_z \mid \mathcal{H}_0 \sim \sigma^2 \chi^2_{2L},
\end{equation}
where $\chi^2_{2L}$ denotes a central chi-square random variable with $2L$ degrees of freedom. 
The chi-square modeling of the test statistic $E_z$ assumes that the
AFDM-domain noise samples over the BD index set $\mathcal{I}_{k,z}$ are
independent and identically distributed. This holds under AWGN and an ideal AFDM transform, for which the additive noise remains white after
affine-domain mapping. Under these conditions, $E_z$ becomes a sum of
independent squared magnitudes, leading to a central chi-square
distribution under $\mathcal{H}_0$ and a non-central chi-square
distribution under $\mathcal{H}_1$.

Under the alternative hypothesis $\mathcal{H}_1$, the $z$-th \ac{BD} actively reflects and
$Y^{\mathrm{Affine}}_{\mathrm{BD}}[k] = S[k] + W[k]$, where $S[k]$ is a deterministic complex term. 
In this case, $E_z$ follows a non-central chi-square distribution with $2L$ degrees of freedom and non-centrality parameter
\begin{equation}
\Lambda_z = \sum_{k \in I_K} |S[k]|^2, 
\qquad
\lambda_z = \Lambda_z / \sigma^2,    
\end{equation}
so that
\begin{equation}
E_z \mid \mathcal{H}_1 \sim \sigma^2 \chi^2_{2L}(\lambda_z),
\end{equation}
where $\chi^2_{2L}(\lambda_z)$ denotes a non-central chi-square random variable with $2L$ degrees of freedom and non-centrality parameter $\lambda_z$. The bit associated with \ac{BD}$_z$ is detected via
\begin{equation}
\hat{b}_z =
\begin{cases}
1, & E_z > \xi, \\
0, & E_z \le \xi,
\end{cases}
\end{equation}
where the threshold $\xi$ is chosen to satisfy a target false-alarm probability $P_{\mathrm{FA}}^\star$. 
Let $F_{\chi^2_{2L}}(\cdot)$ denote the cumulative distribution function (CDF) of the central chi-square distribution with $2L$ degrees of freedom, 
and $F^{-1}_{\chi^2_{2L}}(\cdot)$ denote its inverse CDF (quantile function). 
Then
\begin{equation}
\xi = \sigma^2 F^{-1}_{\chi^2_{2L}}\!\left(1 - P_{\mathrm{FA}}^\star\right)
\end{equation}
ensures $\Pr(E_z > \xi \mid \mathcal{H}_0) = P_{\mathrm{FA}}^\star$. Similarly, let $F_{\chi^2_{2L}(\lambda_z)}(\cdot)$ denote the CDF of the non-central chi-square distribution $\chi^2_{2L}(\lambda_z)$. 
The missed-detection probability for BD$_z$ is given by
\begin{equation}
P_{\mathrm{MD},z}
=
F_{\chi^2_{2L}(\lambda_z)}
\!\left(
F^{-1}_{\chi^2_{2L}}(1 - P_{\mathrm{FA}}^\star)
\right),
\end{equation}
which characterizes the non-coherent detection performance via the non-centrality parameter $\lambda_z$ as a function of the \ac{BD}-reflected energy, the \ac{SNR} and $L$.
\subsection{Coherent detection for the primary OFDM signal}

From (17), after CP removal and $N$-point DFT, the received OFDM block
is expressed in the frequency-domain as
\begin{equation}
\begin{aligned}
Y^{\mathrm{OFDM}}[m]
&= \frac{1}{\sqrt{N}}\sum_{n=0}^{N-1} Y^{\mathrm{OFDM}}[n]\,
e^{-j \frac{2\pi}{N} nm}, \\
&\qquad m = 0,\ldots,N-1
\end{aligned}
\label{coherent_detection}
\end{equation}
Under OFDM demodulation, both the primary channel and the BD-induced
multipath components appear as linear convolutions. Thus, their combined
frequency-domain effect is captured by an effective frequency response
$H_{\text{eff}}[m]$, which is estimated using the same AFDM pilot in frequency domain \cite{BIR_2025}.\\
The coherent OFDM symbol estimate is then obtained via
\begin{equation}
\widehat{S}_{\mathrm{OFDM}}[m]
   = \frac{Y_{\mathrm{OFDM}}[m]}{H_{\text{eff}}[m]},
   \qquad m = 0,\ldots,N-1.
\label{equalization}
\end{equation}
This equalization removes not only the primary BS-\ac{Rx} multipath
distortion, but also the weak backscatter-induced taps superimposed on
the OFDM band. Hence, the primary OFDM data stream is recovered without
interference from BD reflections, while BD detection is performed
independently in the affine domain.

\subsection{Sum-Rate Analysis}
The overall spectral efficiency of the proposed framework is quantified by the sum-rate of the primary signal link and all active \acp{BD} links \cite{mishra2019sum}. Owing to the \ac{OFDM}-\ac{AFDM} domain orthogonality discussed in~(\ref{sec:proposed_scheme}), the primary \ac{OFDM} resources remain fully available for primary link communication, and the \ac{BD} transmissions contribute additional total throughput without sacrificing primary rate.

For the $z^{\text{th}}$\ac{BD}, the effective \ac{SNR} in the affine domain is defined as $\frac{\mathbb{E}\left[ \left| S^{\mathrm{}}_{\mathrm{BD,z}}[k] \right|^2 \right]} {\sigma^2}$.
The corresponding achievable rate is
\begin{equation}
    R_{BD,z} = W \log_2 \bigl(1 + \frac{\mathbb{E}\left[ \left| S^{\mathrm{}}_{\mathrm{BD,z}}[k] \right|^2 \right]} {\sigma^2} \bigr),
\end{equation}
where $S_z[k]$ represents the deterministic, noise-free affine-domain component corresponding to the $z$-th BD, in which the total backscatter throughput is expressed as $R_{\mathrm{BD}} = \sum_{z=1}^{Z} R_{BD,z}$.

For the primary \ac{OFDM} transmission, the direct-link \ac{SNR} after frequency-domain equalization is given as $\mathrm{\gamma}_{\mathrm{p}} = \frac{\mathbb{E}\left[ \left| \hat{S}_{\mathrm{OFDM}}[m] \right|^2 \right]}{\sigma^2}$.
The achievable rate of the primary communication link is then given as
\begin{equation}
    R_{\mathrm{p}} = W \log_2 \bigl(1 + \mathrm{\gamma}_{\mathrm{p}} \bigr).
\end{equation}
Combining both contributions, the system sum-rate is expressed as $R_{\mathrm{sum}} = R_{\mathrm{p}} + R_{\mathrm{BD}}$,
which captures the joint throughput of the primary \ac{OFDM} link and all coexisting \ac{BD} links under the proposed \ac{OFDM}-\ac{AFDM} unified waveform design.

\subsection{Root mean squared error (RMSE) analysis of the estimated range} 
\subsubsection{\ac{BD} sensing}
At the beginning of the sensing stage, all \ac{BD}s remain in a non-reflecting state, enabling the \ac{Rx} to perform bistatic delay probing to identify the maximum excess delay $\ell_{d\max}$ of the propagation environment. 

This bistatic probing assumes that the \ac{BS} and the \ac{Rx} share a common timing reference during the initial environmental sensing phase, consistent with the bistatic AFDM-ISAC architecture in \cite{zhu2024afdm}, where the transmitter and \ac{Rx} operate over a shared AFDM sensing frame. Once these delays are obtained, the BS performs
monostatic sensing and BD delay planning using the same timing
reference.

After this environmental acquisition phase, the \ac{BD}s sequentially switch to reflecting state. During this interval, the BS performs pure monostatic AFDM sensing, allowing it to measure each BD’s round-trip propagation delay. These measurements yield the \ac{BD}’s intrinsic physical delay prior to protocol-controlled manipulation. Once these primary delays are identified, each \ac{BD} introduces an intentional delay~$\ell_{\mathrm{BD},z}$ that is chosen to be strictly greater than the maximum excess delay previously estimated at the \ac{Rx}. This ensures that every \ac{BD} is shifted beyond the environmental delay spread and, consequently, that all \ac{BD}s occupy distinct affine-domain clusters during the communication stage. Such delay planning guarantees non-overlapping \ac{BD} signatures and enables interference-free backscatter detection within the unified OFDM-AFDM symbiotic radio framework.

In the monostatic ISABC architecture, the AFDM probe is a linearly frequency-modulated (LFM) chirp \cite{bemani2024integrated}. Because the BDs remain quasi-static over the sensing interval, the corresponding sensing channel contains delay components only. The continuous-time sensing channel is modeled as

\begin{equation}
h_s(t,\tau)=\sum_{z=1}^{Z} \alpha\,\delta(\tau-\tau_z),
\label{eq:hs_theory}
\end{equation}
where $\tau_z$ is its round-trip propagation delay. The received sensing signal is
\begin{equation}
y_s(t)=\sum_{z=1}^{Z} \alpha\, s_p(t-\tau_z)+w(t),
\label{eq:ys_theory}
\end{equation}
with $w(t)\sim\mathcal{CN}(0,\sigma^2)$ denoting thermal noise. In monostatic sensing, each delay maps directly to the physical range
\begin{equation}
r'_z = \frac{c\,\tau_z}{2},
\label{eq:Rb_theory}
\end{equation}
where $c$ is the speed of light. Thus, estimating $\tau_z$ is sufficient for estimating the range of \ac{BD}.

Dechirping is performed by correlating the received signal with the conjugate of the transmitted pilot $P^\ast_{time}(t)$
\begin{equation}
y_{\mathrm{dech}}(t)=y_s(t)P_{time}^*(t)
=
\sum_{z=1}^{Z} \alpha\,e^{-j2\pi{R_t}\,\tau_z t}+\tilde{w}(t),
\label{eq:dech_theory}
\end{equation}
where $R_t$ is the AFDM chirp rate and $\tilde{w}(t)$ is the dechirped noise. The dechirped 
signal is a superposition of complex exponentials, each with frequency proportional to $\tau_z$. 
Thus, delay estimation reduces to frequency estimation. Applying a Fourier transform yields
\begin{equation}
Y_{\mathrm{dech}}(f)
=
\sum_{z=1}^{Z} \alpha\,\delta(f - {R_t}\tau_z) + W(f),
\label{eq:FFT_theory}
\end{equation}
so each BD produces a peak at frequency $f_z={R_t}\tau_z$, from which the delay estimate follows
\begin{equation}
\hat{\tau}_z=\frac{f_z}{{R_t}},
\qquad
\hat{r}_z=\frac{c\,\hat{\tau}_z}{2}.
\label{eq:tau_range_hat_theory}
\end{equation}

\subsubsection{RMSE of range estimation}
Let $r_z$ denote the true bistatic range associated with \ac{BD}$_z$ and $\widehat{r}_z$ its estimate obtained from the \ac{AFDM}-domain delay of the reflected pilot. The range estimation error is defined as
\begin{equation}
\varepsilon_{r,z} = \widehat{r}_z - r_z.
\end{equation}
The quality of the estimator is characterized by the \ac{RMSE}, given by
\begin{equation}
\mathrm{RMSE}_r = \sqrt{\mathbb{E}\big[\varepsilon_{r,z}^2\big]}~,
\end{equation}
where the expectation is taken over the randomness of the noise and the fading channel affecting the \ac{BD} link. In the considered system, $\widehat{r}_z$ is obtained by first estimating the
propagation delay $\widehat{\tau}_z$ from the \ac{AFDM}-domain pilot shift and then mapping it to range via
\begin{equation}
\widehat{r}_z = c\,\widehat{\tau}_z,
\end{equation}
the corresponding delay error is $\varepsilon_{\tau,p} = \widehat{\tau}_z - \tau_z$ and range error are related by
\begin{equation}
\varepsilon_{r,z} = c\,\varepsilon_{\tau,z},
\end{equation}
so that
\begin{equation}
\mathrm{RMSE}_r = c\,\mathrm{RMSE}_\tau,
\end{equation}
where $\mathrm{RMSE}_\tau = \sqrt{\mathbb{E}\big[\varepsilon_{\tau,z}^2\big]}$
denotes the \ac{RMSE} of the delay estimate. This expression provides theoretical performance metric that depends on the \ac{AFDM} waveform parameters, the BD reflection coefficient, and the
operating \ac{SNR}.

\subsection{Computational complexity}
The computational cost of the proposed architecture is dominated by the \ac{BS} synthesis of the unified transmit block and by low-rate \ac{BD} delay switching. Equation (\ref{eqn:comb}) shows that, per block, the \ac{BS} generates an $s^{\mathrm{OFDM}}[n]$ through an $N$-point IDFT and $P_{time}[n]$ pilot through an $N$-point IDAFT to form the composite transmit waveform 
$s_{p}[n]$. Since both \ac{OFDM} and \ac{AFDM} are realized via inverse transform operations of \ac{FFT} order, the \ac{BS} side generation scales as $\mathcal{O}(N\log N)$ for \ac{OFDM} plus $O(N\log N)$ for \ac{AFDM} \cite{ahmad2025joint}, with an additional linear $\mathcal{O}(N)$ cost for superposition. Hence, the per-block transmit side complexity is
\begin{align}
    C_{\mathrm{BS}} &= \mathcal{O}(N\log N) + \mathcal{O}(N\log N) + \mathcal{O}(N)\nonumber\\
    &
    = \mathcal{O}(2N\log N + N).
\end{align}

On the device side, each \ac{BD} is a passive \ac{SAW}/\ac{BAW} based delay-line node in which a reflection-control switch selects one of the available delay paths per bit; a `0' corresponds to absorption and a `1' triggers a deterministic delayed reflection. This operation requires no digital baseband processing or active \ac{RF} generation, implying constant per-device effort. Therefore, for $Z$ simultaneous \acp{BD}, the aggregate device-side overhead grows only linearly as
\begin{equation}
    C_{\mathrm{BD}} = \mathcal{O}(J).
\end{equation}
Finally, the overall system complexity per unified \ac{OFDM}-\ac{AFDM} block is
\begin{equation}
    C_{\mathrm{sys}} = \mathcal{O}(2N\log N + N + J),
\end{equation}
which preserves standard \ac{OFDM} FFT-order scaling while enabling multi-\ac{BD} backscatter communication and sensing with only low-order linear growth in the number of devices. Unlike conventional ambient or symbiotic backscatter systems, the proposed architecture does not require iterative interference cancellation or correlation-based maximum-likelihood (ML) detection to separate weak BD signals from a strong direct link, making their complexity grow as $\mathcal{O}(Z N^2)$ with $N$ and $Z$.
\subsection{Power Consumption}
The proposed unified OFDM-AFDM signaling framework enables the BS to transmit primary communication, multi-BD backscatter detection, and monostatic sensing under a single transmit-power budget. The OFDM data symbols and the AFDM pilot share the same RF resources. Therefore, the allocation of power between these components fundamentally determines the performance of all three functions. To characterize this relationship, let $E_{\mathrm{pilot}}$ and $E_{\mathrm{data,eff}}$ denote the AFDM pilot energy and the effective OFDM data energy per block, respectively, where $P_{\mathrm{data}}$ is the power allocated to data. Their ratio is expressed as

\begin{equation}
\eta = 10 \log_{10}\!\left(\frac{E_{\mathrm{pilot}}}{E_{\mathrm{data,eff}}}\right)
      = 10 \log_{10}\!\left(\frac{P_{pilot}}{P_{\mathrm{data}}}\right).
\end{equation}
which directly reflects the relative SNR levels available for sensing ($\gamma_{\mathrm{RMSE}}$) and BD detection ($\gamma_{\mathrm{PMD}}$).
For monostatic sensing, the AFDM pilot acts as a deterministic probing waveform whose matched-filter output exhibits a processing gain proportional to its energy. Consequently, the effective sensing SNR is given by
\begin{equation}
    \gamma_{\mathrm{RMSE}} = \frac{E_{\mathrm{pilot}}}{N_{0}}.
    \label{eq:gamma_rmse}
\end{equation}
Where, $N_0$ denotes the single-sided noise power spectral density. The variance of the delay estimate decreases inversely with $E_{\mathrm{pilot}}$. Hence, increasing $P_{\mathrm{pilot}}$ directly enhances the resolution and robustness of delay estimation.

For ambient backscatter detection at the Rx, the detectable \ac{BD} signature depends on both the amplitude of the \ac{BD} reflection and the strength of the \ac{AFDM} pilot, while the reflected OFDM component scales with $E_{\mathrm{data,eff}}$. Thus, reliable \ac{BD} detection requires an appropriate balance between pilot and data powers.

\section{Simulation results}
\label{sec:sim_results}
\begin{figure*}[t!]
    \centering
    \begin{minipage}[t]{0.33\textwidth}
        \centering
        \includegraphics[width=\linewidth]{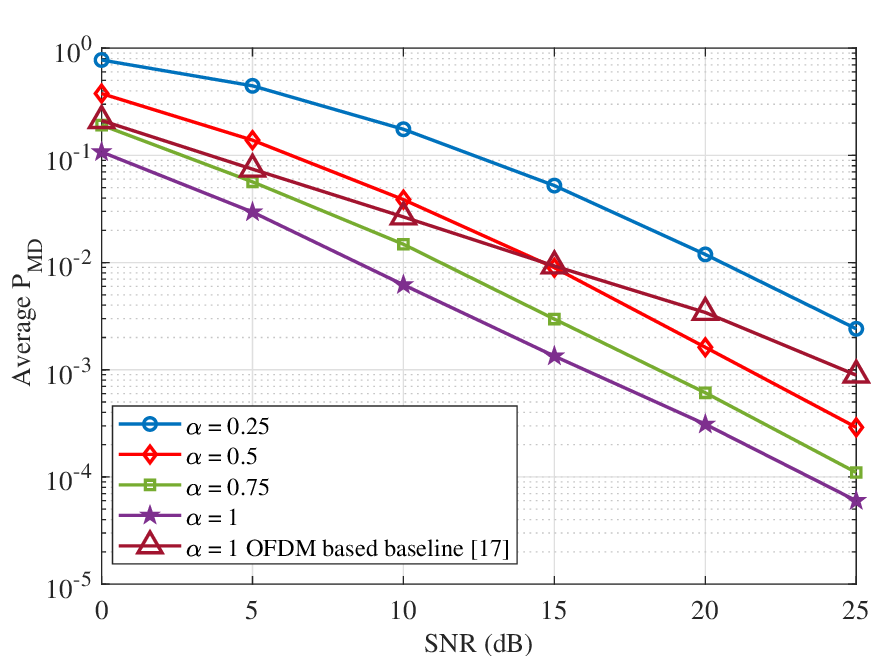}
        \caption{The PMD of the backscatter communication vs. SNR with different reflection coefficient of non-coherent detector for the proposed scheme with N=256, $c_1'$=8, $P_{\mathrm{pilot}}$= 21.1 dB, and \ac{BD} = 3.}
        \label{fig:pmd1}   
    \end{minipage}
    \hfill
    \begin{minipage}[t]{0.32\textwidth}
        \centering
        \includegraphics[width=\linewidth]{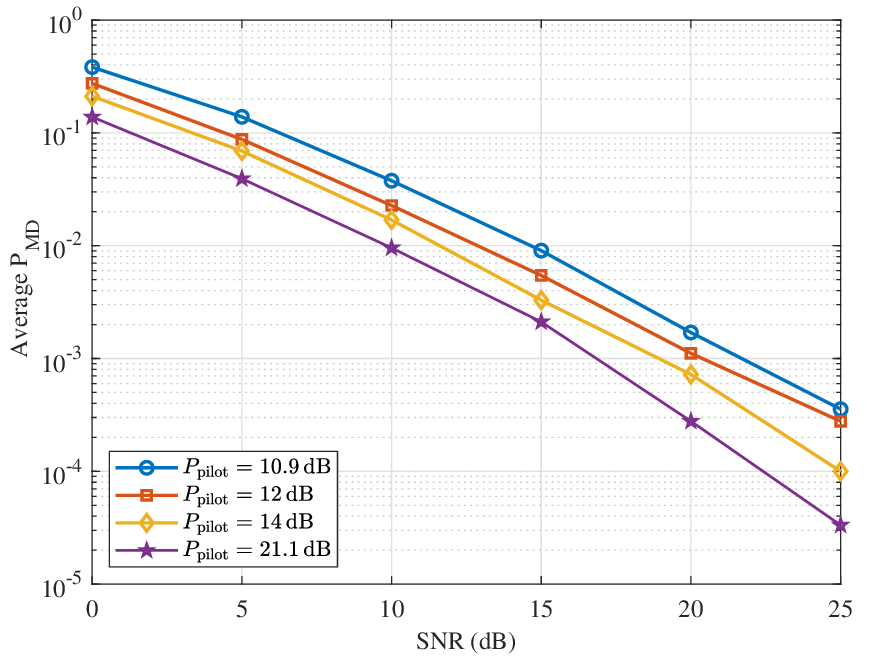}
        \caption{The PMD of the backscatter communication vs. SNR with different pilot power of non-coherent detector for the proposed scheme for $\alpha$ = 1 and N=256, $c_1'$=8, and \ac{BD} = 3.}
        \label{fig:pmd2}
    \end{minipage}
    \hfill
    \begin{minipage}[t]{0.32\textwidth}
        \centering
        \includegraphics[width=\linewidth]{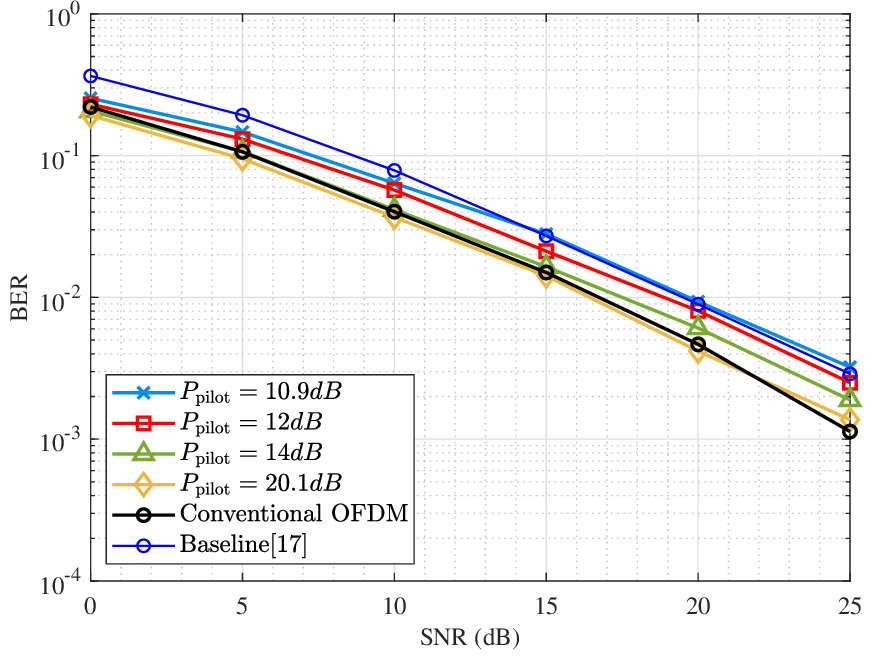}
        \caption{The BER of the primary communication vs. SNR for the proposed scheme with N=256, $c_1'$= 8, and \ac{BD} = 3.}
        \label{fig:imageber}
    \end{minipage}
\end{figure*}
In this section, we provide a numerical performance evaluation of the proposed framework, jointly assessing its communication and sensing capabilities. In particular, we investigate the impact of key system parameters on the reliability of both the \ac{BC} and primary communication links, as well as on the \ac{BS}'s sensing accuracy. Unless otherwise specified, the simulation parameters setup are listed in Table~\ref{tab:sim_params}. The obtained results are benchmarked against the baseline scheme in \cite{uledi2025interference} in terms of detection accuracy, i.e., probability of missed detection. In this benchmark method, the backscatter tag applies a predefined frequency shift keying (FSK) such that its response is relocated to an empty subcarrier in the \ac{OFDM} spectrum. The receiver then performs energy detection on that vacant subcarrier; if a significant energy rise is observed at the shifted location, the \ac{BD} is declared active.

\begin{table}[!ht]
    \centering
    \caption{Simulation Parameters}
    \label{tab:sim_params}
    \setlength{\tabcolsep}{4pt} 
    \renewcommand{\arraystretch}{1} 
    \begin{tabular*}{\columnwidth}{@{\extracolsep{\fill}}lll}
        \hline
        \textbf{Parameter} & \textbf{Symbol} & \textbf{Value} \\
        \hline
        Modulation order              & --       & 4-QAM \\
        \ac{DFT} size                 & $N$      & $256$ \\
        Affine-domain index           & $i$         & $1$ \\
        Cyclic prefix length          & $C_p$    & $N/4$ \\
        AFDM spreading factor           & $c_1'$    & $8$ \\
        \ac{BD} reflection coefficient    & $\alpha$ & $1$ \\
        Target false-alarm probability& $P_{\mathrm{FA}}$ & $10^{-3}$ \\
        
        \hline
    \end{tabular*}
\end{table}

\subsection{Error rate}
\figurename~\ref{fig:pmd1} presents average $P_{\mathrm{MD}}$ of the backscatter link for several values of the reflection coefficient~$\alpha$. As expected, $P_{\mathrm{MD}}$ decreases monotonically with SNR, with the lowest detection errors obtained when $\alpha = 1$. For example, at $\mathrm{SNR}=25$~$\mathrm{dB}$, increasing $\alpha$ from $0.25$ to $1$ reduces the average $P_{\mathrm{MD}}$ from approximately $10^{-2}$ to $10^{-4}$. This highlights the inherent fragility of weak reflections, since for small~$\alpha$ values the backscattered component is of the same order as the noise variance.
What distinguishes the proposed AFDM-based detector from the classical ambient OFDM behavior of chirp waveforms under energy detection is that AFDM employs quadratic-phase chirps whose energy is uniformly distributed over time. After de-chirping at the \ac{Rx}, the chirp energy collapses into a single sharp, non-fading peak in the affine domain, producing a highly stable and high-contrast signature that is exceptionally well-suited for non-coherent energy detection.
In contrast, the OFDM-based baseline \cite{uledi2025interference} detector operates on a waveform whose instantaneous amplitude fluctuates significantly due to constructive and destructive subcarrier interference, resulting in a diffuse detection statistic. Consequently, the baseline method loses precision in the regime where low-power tags operate, whereas the proposed AFDM approach enables nearly two orders of magnitude improvement in $P_{\mathrm{MD}}$ at moderate SNR for $\alpha = 1$, while still maintaining strong detection capability even for $\alpha = 0.25$.
\subsection{Sum-rate}
\begin{figure*}[t]
    \centering
    \begin{minipage}[t]{0.45\linewidth}
        \centering
        \includegraphics[width=\linewidth]{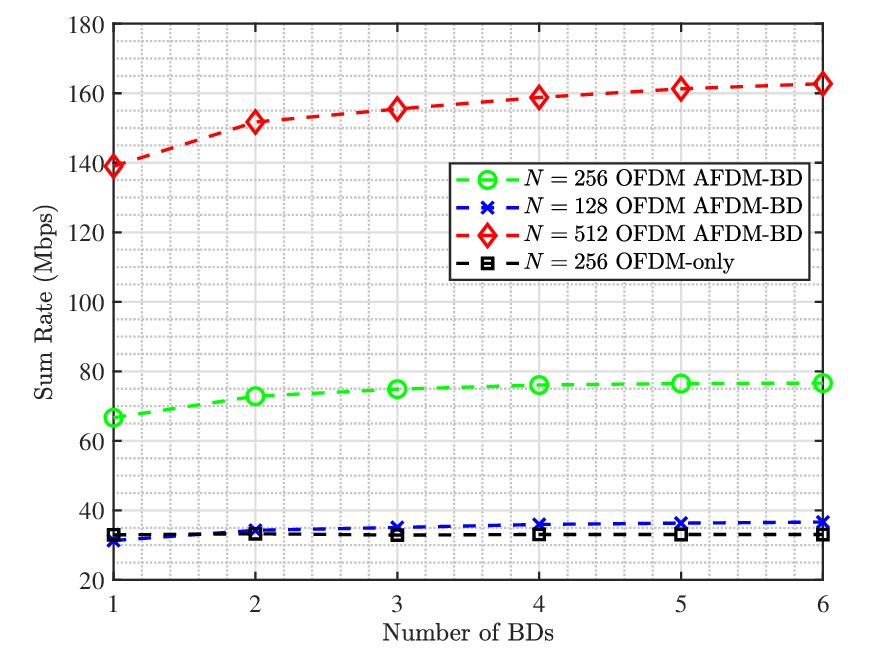}
        \caption{The sum-rate performance vs. number of BDs with different number of subcarriers for $\alpha=1$ at SNR = 25 dB.}
        \label{fig:sum1}
    \end{minipage}
    \hfill
    \begin{minipage}[t]{0.45\linewidth}
        \centering
        \includegraphics[width=\linewidth]{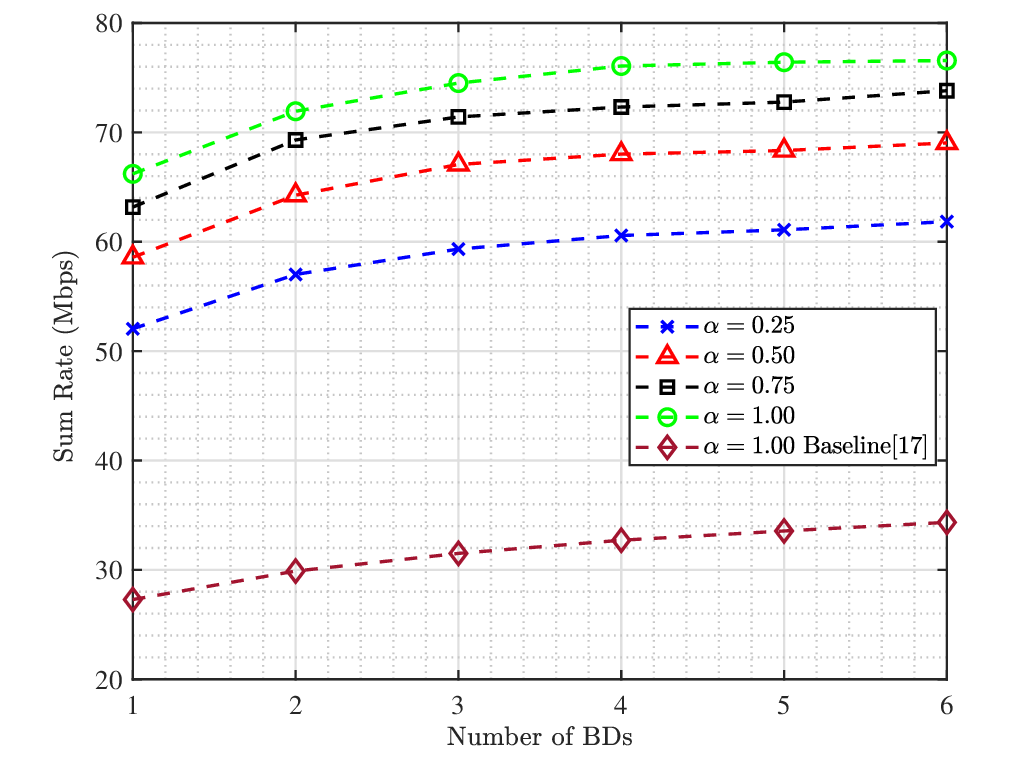}
        \caption{The sum-rate performance vs. number of BDs, with different reflection coefficients at SNR = 25 dB, N=256s, and $c_1'$=8.}
        \label{fig:sum2}
    \end{minipage}
\end{figure*}
\figurename~\ref{fig:pmd2} depicts the impact of the AFDM pilot power on the detection of backscatter signals. As shown, $P_{\mathrm{MD}}$ consistently decreases as the pilot power increases. For instance, at $\mathrm{SNR}=25$~dB, increasing the pilot power from $10.9$~dB to $21$~dB reduces the average $P_{\mathrm{MD}}$ from values in the range of $10^{-3}$ to $10^{-4}$. This confirms that appropriate pilot power allocation is essential for reinforcing weak backscatter reflections, particularly in low- and medium-SNR regimes. Within the proposed AFDM framework, the chirp's stable, non-fading energy structure makes it inherently well-suited for low-power non-coherent detection, enabling substantial improvements in $P_{\mathrm{MD}}$.

\figurename~\ref{fig:imageber} examines the influence of the AFDM pilot power on the BER of the primary OFDM communication link across different SNR values. A slight improvement in BER is observed as the pilot power $P_{\mathrm{pilot}}$ increases. For example, at $\mathrm{SNR}=25$~dB, raising the pilot level from $10.9$~dB to $21$~dB reduces the BER from the order of $10^{-2}$ to values closer to $10^{-3}$. This behavior highlights a key property of the proposed OFDM-AFDM coexistence framework. The AFDM backscatter operation is fully contained within the CP region, and the resulting chirp-based reflections do not distort the orthogonality of the OFDM data symbols.
As a result, the primary OFDM subcarriers remain unaffected, and the obtained BER matches that of a conventional OFDM system operating without AFDM or backscatter activity. A second key observation emerges when comparing these curves to the baseline FSK-based scheme \cite{uledi2025interference}. Across the entire SNR range, the Baseline curve consistently lies above both the Conventional OFDM and the proposed OFDM-AFDM curves, indicating notably worse BER performance. This degradation is expected, as it is sensitive to small frequency shifts. Therefore, it verifies that the proposed system achieves true interference-free coexistence, supporting low-power backscatter communication while maintaining the native performance of the OFDM link.

\figurename~\ref{fig:sum1} demonstrates the influence of the OFDM-AFDM block size $N$ on the overall sum-rate performance of the system as the number of backscatter devices increases. In this coexistence setting, the overall throughput reflects not only the sustained OFDM data rate but also the additional contribution introduced by the BD reflections, which naturally elevates the sum-rate as more devices become active. The results indicate that the system throughput increases with larger $N$, primarily because a higher subcarrier count is associated with a longer CP, which in turn provides a larger interference-free region for AFDM-based backscatter embedding. Consequently, systems with larger $N$ can accommodate more BDs without mutual overlap in the affine domain.
For example, while the exact numerical values depend on the specific parameter configuration, the general trend consistently shows that a larger block size (e.g., $N = 512$) achieves a significantly higher sum-rate than a smaller block size (e.g., $N = 128$) when operating with the same number of \ac{BD}s. Specifically, when 6 BDs are active, the configuration with 
$N = 128$ yields a sum-rate of approximately $39~\text{Mbps}$, 
whereas the $N = 256$ and $N = 512$ settings achieve around 
$78~\text{Mbps}$ and $160~\text{Mbps}$, respectively, which is 
even larger than the conventional OFDM-only system that 
achieves about $31~\text{Mbps}$ with $256$ subcarriers.
This improvement arises because a larger \ac{AFDM} grid offers finer delay resolution, allowing each \ac{BD}’s chirp-induced delay signature to occupy a distinct \ac{CP} region. The chirp structure ensures that each \ac{BD} produces a stable, well-localized affine-domain peak, enabling efficient multi-device separation and symbol recovery. Overall, it confirms that increasing $N$ enhances the system’s capacity to support multiple low-power \ac{BD}s simultaneously, while maintaining reliable detection and high aggregate throughput.

\figurename~\ref{fig:sum2} illustrates the impact of the backscattered power, captured by the reflection coefficient $\alpha$, on the sum-rate performance of the system as the number of BDs increases. It is observed that higher values of $\alpha$ consistently result in better performance, since stronger backscattered components contribute more reliably to the received signal at the \ac{Rx}. For any fixed number of BDs, increasing $\alpha$ leads to a noticeable increase in the achievable sum-rate, whereas small $\alpha$ values yield reduced throughput due to weaker and more noise-sensitive reflections. For example, at 6 BDs, the system achieves approximately $78$~Mbps for $\alpha = 1$, while $\alpha = 0.25$ yields around $63$~Mbps, confirming the substantial gain provided by stronger reflections. In comparison, the baseline OFDM detection approach \cite{uledi2025interference}, which assigns an empty subcarrier per BD to detect frequency shifts, achieves only about 35 Mbps at 6 BDs, highlighting the much lower spectral efficiency of the conventional design. Although the numerical values depend on the specific simulation parameters, the overall trend is that larger $\alpha$ enhances the effective SNR of each BD link, thereby improving both detection reliability and the aggregate rate. This behavior is further reinforced by the AFDM structure, where each BD’s reflection produces a chirp-based delay signature that collapses into a sharp affine-domain peak after de-chirping. As $\alpha$ increases, these peaks become more pronounced and easier to separate, enabling the system to support multiple low-power BDs while still achieving a steadily increasing sum-rate.
\begin{figure*}[!t]
    \centering

    \begin{minipage}[t]{0.32\linewidth}
        \centering
        \includegraphics[width=\linewidth]{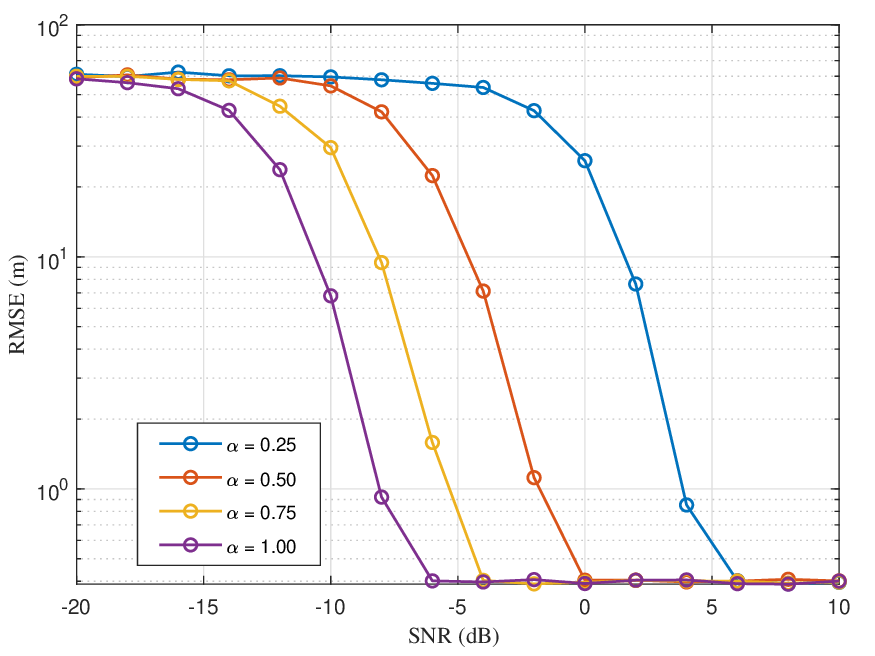}
        \caption{RMSE performance vs. SNR ($\gamma_{\rm RMSE}$) for different reflection coefficients.}
        \label{fig:rmse2}
    \end{minipage}
    \hfill
    \begin{minipage}[t]{0.31\linewidth}
        \centering
        \includegraphics[width=\linewidth]{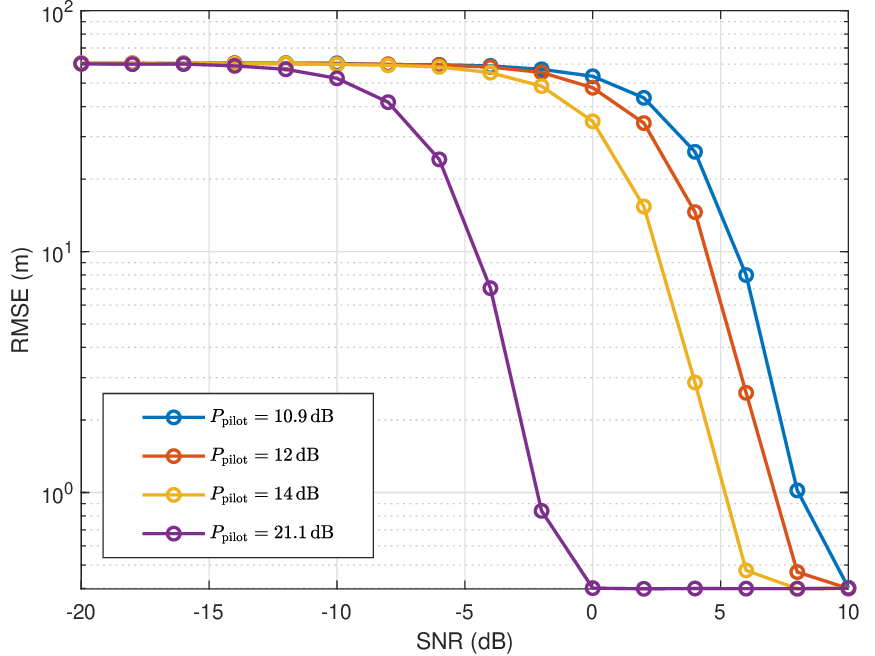}
        \caption{RMSE performance vs. SNR with different $P_{\mathrm{pilot}}$ at $\alpha=1$.}
        \label{fig:rmse1}
    \end{minipage}
    \hfill
    \begin{minipage}[t]{0.33\linewidth}
        \centering
        \includegraphics[width=\linewidth]{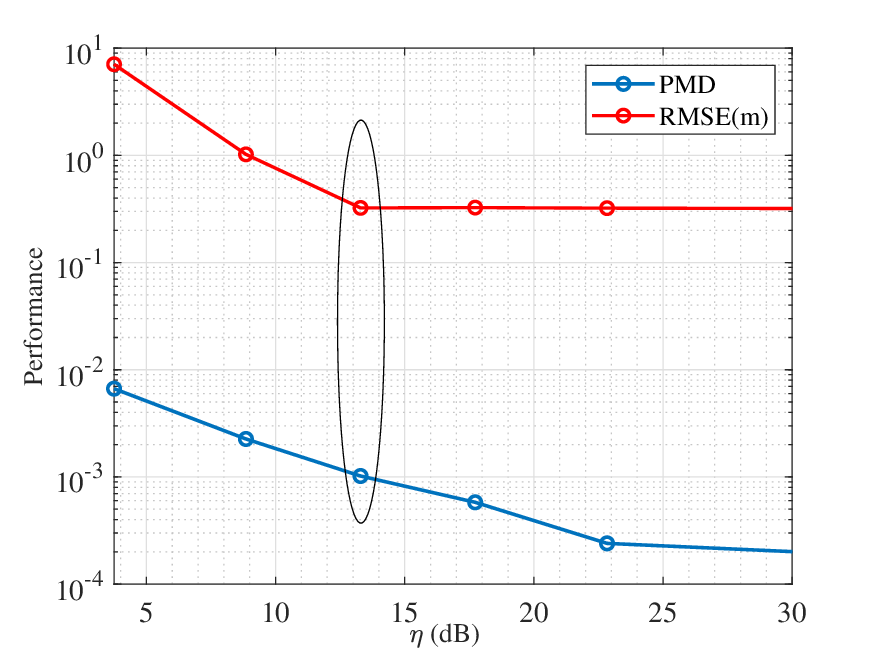}
        \caption{ISABC performance vs. power ratio $\eta$.}
        \label{perfpow}
    \end{minipage}

\end{figure*}

\subsection{RMSE}
\figurename~\ref{fig:rmse2} illustrates the impact of the backscattering strength~$\alpha$ on the RMSE of the bistatic range estimate as a function of SNR. As the SNR increases, each curve exhibits a sharp transition in which the RMSE rapidly drops to values close to zero, indicating highly accurate range estimation. Larger $\alpha$ values shift this collapse point toward much lower SNRs and yield uniformly smaller RMSE across the entire SNR range, demonstrating that stronger backscatter reflections substantially enhance sensing robustness in noise-limited regimes. For example, the RMSE collapse occurs near $-10$~dB when $\alpha = 1$, whereas the same transition appears only after approximately $+5$~\text{dB} when $\alpha = 0.25$. This highlights the sensitivity of the collapse threshold to the reflection strength.
The key result is that even weak reflections ultimately achieve the same asymptotic RMSE once the operating SNR becomes sufficiently large, with all curves converging below $0.5$~m for SNR values above $10$~dB. This again confirms the inherent robustness of the AFDM sensing mechanism. Although stronger reflections improve performance under low-SNR conditions, accurate delay estimation remains attainable even for small~$\alpha$, provided that the SNR is moderate. Thus, the proposed system can reliably support a wide range of BD reflection characteristics without imposing strict hardware constraints.

\figurename~\ref{fig:rmse1} shows the impact of the pilot power $P_{\mathrm{pilot}}$ on the RMSE of the monostatic range estimate. When $\alpha = 1$, higher pilot powers consistently yield improved sensing accuracy across all SNR levels. As \ac{SNR} increases, \ac{RMSE} drops sharply, reaching very small values at moderate \ac{SNR}, indicating a rapid improvement in sensing accuracy. In addition, higher $P_\text{pilot}$ powers shift this transition to lower \ac{SNR} indicating a consistent lower \ac{RMSE}, highlighting the critical role of sufficient pilot energy for reliable sensing in noisy conditions. A key observation is that the system maintains accurate range estimation of the weak backscatter signal even when the pilot power is significantly reduced. For example, with only $10.9$~dB of pilot energy, a level that generates a very weak de-chirped peak, the proposed \ac{BD} sensing scheme still achieves sub-meter accuracy once the SNR exceeds a moderate level approximately $0$-$2$~dB. These observations confirm the critical role of proper pilot-power allocation in ensuring robust sensing within OFDM-AFDM coexistence frameworks.
\subsection{Performance vs. power ratio}
\figurename~\ref{perfpow} illustrates the ISABC performance as a function of the power ratio $\eta$ showing both the PMD and the sensing \ac{RMSE} under a fixed per-block transmit-power budget. Varying $\eta$ thus represents a reallocation of the same total energy between \ac{AFDM} and \ac{OFDM} data. In the low-$\eta$ regime, pilot energy is insufficient, leading to a high PMD and poor ranging accuracy. As $\eta$ increases to a moderate level, both metrics improve rapidly, indicating that a modest pilot-energy boost simultaneously strengthens \ac{BC} detection and sensing precision to $10^{-3}$ and $0.33 m$ respectively. Beyond the apparent knee $\eta$ = 13.5 dB, the curves saturate, so additional pilot power provides only marginal performances gains. Hence, the knee region identifies the most energy-efficient operating point, where the unified waveform framework design meets sensing and \ac{BC} requirements with minimal pilot consumption power and maximal retained data efficiency.

\section{CONCLUSION}
\label{sec:concl}
This study introduces a low-power \ac{ISABC} architecture that exploits \ac{OFDM}-\ac{AFDM} waveform orthogonality to realize interference-free \ac{BC} and sensing. By placing sparse \ac{AFDM} pilots and enforcing deterministic affine-domain delay shifts within the clean \ac{CP}, the proposed design suppresses both \ac{DLI} and \ac{IBDI} without degrading the primary \ac{OFDM} link performance. The \ac{AFDM} chirp structure further enables accurate monostatic ranging of multiple \acp{BD} at the \ac{BS}, by providing well-localized delay signatures that remain separable under the proposed affine-domain shifting. The results show that both the PMD and range \ac{RMSE} consistently improve as the \ac{BD} reflection coefficient and the pilot power increase, whereas the primary communication link maintains a BER performance. Importantly, these gains are achieved with efficient computational complexity for waveform generation and for the \acp{BD} operations, and only linear growth with the number of active \acp{BD}, hence supporting scalable dense \ac{IoT} deployments. The framework therefore offers an effective reliability sensing power trade-off under efficient spectral reuse, and can be directly adapted to practical multi-carrier systems such as Wi-Fi, LTE, and 5G. Future work will focus on robust separation of channel and \ac{BD} induced delays in rich multipath, extending the model to delay–Doppler sensing under mobility and hardware impairments.
\bibliographystyle{IEEEtran}
\bibliography{IEEEfull,references}
\end{document}